# AARC: First draft of the Blueprint Architecture for Authentication and Authorisation Infrastructures

A. Biancini[12], L. Florio[1], M. Haase[2], M. Hardt[3], M. Jankowsi[13], J. Jensen[4], C. Kanellopoulos[5], N. Liampotis[5], S. Licehammer[6], S. Memon[7], N. van Dijk[8], S. Paetow[9], M Prochazka[6], M. Sallé[10], P. Solagna[11], U. Stevanovic[3], D. Vaghetti[12]

[1] GEANT Association (GEANT), Amsterdam, Netherlands
[2] DAASI International GmbH (DAASI), Tübingen, Germany
[3] Karlsruhe Institute of Technology (KIT), Karlsruhe, Germany
[4] Science and Technology Facilities Council (STFC), Oxfordshire, United Kingdom
[5] Greek Research and Technology Network (GRNET), Athens, Greece
[6] Zajmove Sdruzeni Pravnickych Osob (CESNET), Prage/Brno, Czech Republic
[7] Forschungszentrum Jülich GmbH (FZJ), Jülich, Germany
[8] SURFnet B.V., Utrecht, Netherlands
[9] The JNT Association (JANET), Oxford, United Kingdom
[10] Nikhef National Institute for Subatomic Physics, Amsterdam, Netherlands
[11] European Grid Initiative (EGI) Amsterdam, Netherlands
[12] Consortium GARR (GARR), Italy
[13] Instytut Chemii Bioorganicznej Pan (PSNC), Poznan, Poland

# Executive Summary

This document presents the first draft version of the AARC Blueprint Architecture. This first version of the blueprint architecture provides a reference architecture to *help e-infrastructure operators and technical architects in various research communities to design secure, scalable, user-friendly, and interoperable AAIs*.

The document has been circulated for comments among several research and e-infrastructures (such as Elixir, Dariah or CERN) that are potential implementers of the proposed architecture. The updated version of the document, which includes the received inputs, was published as the milestone document MJRA1.4[1], due at the end of July 2016. Any feedback to this document will help AARC shape the next versions of the AARC Blueprint Architecture.

The content of this document was driven by the user communities and e-infrastructure*s* requirements identified by the AARC architecture team, and it addresses the challenges that inhibit a wide adoption of federated access for eScience.

The document is organised as follows: Section 2 will present the general architectural landscape as it appears to date. This leads to proposing a general Blueprint Architecture, which will be detailed in Section 3.

In Section 4 we discuss the proposed architecture and analyse how it can match the identified requirements[2]. Two appendices are provided that summarize the collected requirements (A) and the set of AAI Architectures analysed (B).

---

[1] https://aarc-project.eu/wp-content/uploads/2016/08/MJRA1.4-First-Draft-of-the-Blueprint-Architecture.pdf
[2] https://aarc-project.eu/wp-content/uploads/2015/10/AARC-DJRA1.1.pdf

The acronyms used are those defined in the AARC document in Terms and Definitions[3]

# 1. Introduction

This document presents the first draft version of the AARC Blueprint Architecture. Since the publication of the draft version in may 2016, the team has been presenting it and gathering feedback, to present it to the e-infrastructures, research infrastructures, and research collaborations to solicit feedback on the proposed approach. Any feedback to this document will be taken into account in shaping the next versions of the AARC Blueprint Architecture.

The document targets the technical architects and implementers of IT infrastructures for international research collaboration. The goal of the document is to provide a general blueprint AAI architecture that can be used as a reference for the design and implementation of integrated and interoperable AAI solutions in the R&E sector.

In the last five years, eduGAIN has grown to become a global network of R&E identity federations around the world, simplifying access to content and resources for the international R&E communities. eduGAIN, by enabling the trustworthy exchange of information related to identity, authentication, and authorization (AAI) at a global scale, is a solid foundation for federated access in R&E. AAI architects and implementers can build advanced, tailored technical solutions on top of eduGAIN that meet the requirements of their research communities. Although one solution cannot fit all, by defining and adopting common best practices, architectural patterns and interfaces, we can build AAIs that are interoperable and which can be integrated in a wider ecosystem of e-Infrastructures and services for R&E.

In the last year, the AARC project has been working together with e-Infrastructures, research infrastructures, research communities, AAI architects and implementers in order to collectively define a set of architectural building blocks and implementation patterns that will allow the development of interoperable technical solutions for international intra- and inter-disciplinary research collaborations. The goal of AARC is to provide guidelines and building blocks to support the development of interoperable AAIs. The proposed blueprint architecture aims to offer a framework compatible with existing solutions and which would ensure cross-implementation compatibility and interoperability.

Between May and September 2015, we engaged in a discussion with scientific communities, infrastructure and service providers through a series of questionnaires and interviews, in order to capture their use cases, requirements, and existing technical solutions. The requirements captured by the TERENA AAA Study[4] and the FIM4R paper[5] were revisited and updated in light of the new findings and published as an analysis document[6]. This work was complemented by a parallel activity in which we conducted a "market research"[7] for existing software solutions that are used today in the implementation of AAI solutions in the wider R&E sector.

---

[3] https://docs.google.com/document/d/18AllfUKLi90f1odm6hINkQvRljbFhy9lfkY1M447uBQ
[4] https://www.terena.org/publications/files/2012-AAA-Study-report-final.pdf
[5] https://cds.cern.ch/record/1442597/files/CERN-OPEN-2012-006.pdf?version=2
[6] https://aarc-project.eu/wp-content/uploads/2015/10/AARC-DJRA1.1.pdf
[7] https://aarc-project.eu/wp-content/uploads/2016/01/MJRA1.1-Existing-AAI-and-available-technologies.pdf

# 2 AAI landscape for global research collaboration

Federated access is fundamental for the international research collaboration. Researchers need to be able to access services, share data, and collaborate with the members of their scientific communities. Emerging research infrastructures (RI) comprise facilities such as computational clusters, storage, information systems; they are typically governed by community-specific policies that regulate access and publication. RIs typically provide services that are specifically tailored to the specific scientific disciplines.  As a result, most of the RIs implement specific access policies for their scientific communities.

The way researchers collaborate within scientific communities can vary significantly from community to community. On the one hand, there are highly structured communities with thousands of researchers who can be virtually anywhere in the world. Typically, these communities have been working together for a long time. They want to share and have access to a wide range of resources, and they have had to put in place practical solutions to make the collaborations work. On the other hand, one finds a diverse number of smaller research communities working within specific or across scientific disciplines. Typically, these are either nascent communities being established around new scientific domains, or they are communities in specific domains that did not have to promote widespread and close collaboration among the researchers. And of course, in between these two extremes, there are many scientific communities of varying size, structure, history etc.

The growing number of researchers and services with advanced requirements such as community based authorization, support for credential delegation, attribute aggregation from multiple sources, requires additional functional components on top of eduGAIN. This is the conclusion of the Analysis of user- community requirements  deliverable (DJRA1.1[8]).

Based on the  analysis of the existing and developing AAIs, outlined in Appendix B, similar approaches that target these additional functional components were identified.

## 2.1 How research communities can use eduGAIN today

Research communities and e-infrastructures can already rely on eduGAIN and the underlying identity federations to authenticate their users.

The picture below (Figure 1[9]) depicts the standard approach in which different services  become available via eduGAIN through a participating federation (it is not relevant whether a federations is implemented as 'full-mesh' or as 'hub-and-spoke[10];').

---

[8] https://aarc-project.eu/wp-content/uploads/2015/10/AARC-DJRA1.1.pdf
[9] http://services.geant.net/edugain/About_eduGAIN/Pages/Home.aspx
[10] https://wiki.edugain.org/Federation_Architecture

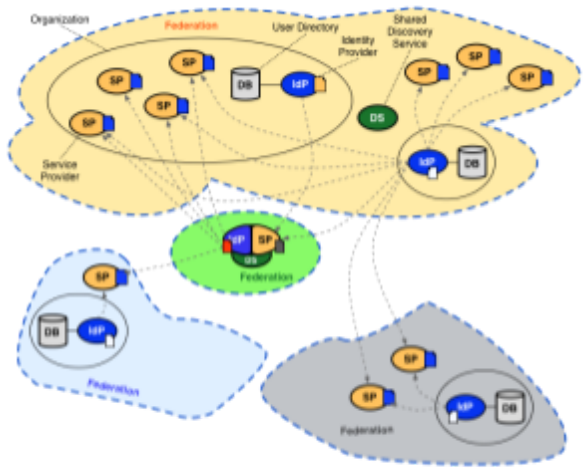 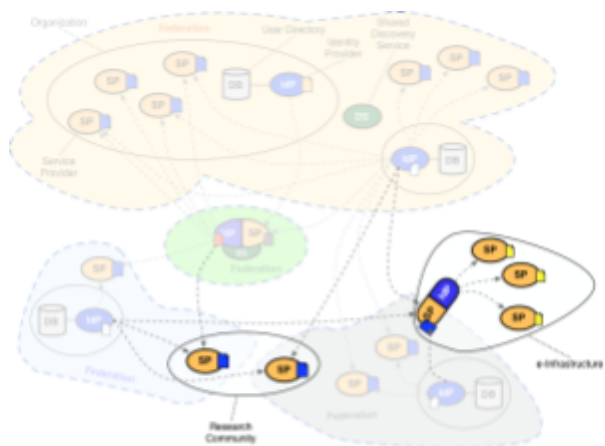

*Figure 1: SAML interfederation, as provided (for example) by eduGAIN - Top part (almond color) shows the example of the "full mesh federation", while the lower part (green, blue, and gray) shows the "hub-and-spoke" federation*

*Figure 2: Examples of a research community and an e-Infrastructure in eduGAIN*

Figure 2 depicts two separate cases of research communities, which use eduGAIN for providing federated access to their services. In left of the lower half of the diagram two service providers, belonging to the same research community, are depicted. This research community has set up their own SPs inside different federations, so that federated members can log in.

Setting up a new SP into which all users of a community can login, is however, generally perceived to be a complex and time consuming task. E.g. the coordination with federation and identity provider operators in order to permit the release of the necessary attributes is time consuming process with very slow results. This is due to various reasons, such as differences in national privacy laws, attribute release policies and the diverse representations of users delivered by different IdPs.

The second scenario presented on the right side of figure 2 represents a research or e-Infrastructure, which is connected to eduGAIN via a single SP, acting like an SP-IdP proxy. The SP-IdP proxy component augments the authenticator and attributes that come from the 'conventional R&E IdPs' with elements that are essential for the infrastructure or research community: a persistent non-reassigned identifier, assurance level, maybe add community membership roles and groups, maybe add reputation. In that, the RIs shield themselves from the heterogeneity of the global R&E federation space, and they make it easier for themselves since they can now 'hide' all their services behind just a single proxy..

# 3 A Blueprint Architecture

The existing federated AAI infrastructure is presented in Figure 3. The diagram illustrates the existing workflows and visualises flows between the components. The level of detail is intentionally low so that all relevant components can be included.

This diagram (figure 3) defines four layers, namely: **User Identities, Attribute Enrichment, Translation,** and **End Services**. Each layer contains one or more components. The diagram does not strictly depict deployment scenarios. SPs are likely situated in a different organisational domain and nation than other services or users. Identities are possibly provided by services different from those that provide attributes.

In order to keep the architectural drawing simple and clear, the diagram intentionally excludes **user-workflows** (especially details of redirections) and required **trust models** since they are out of scope in this document.

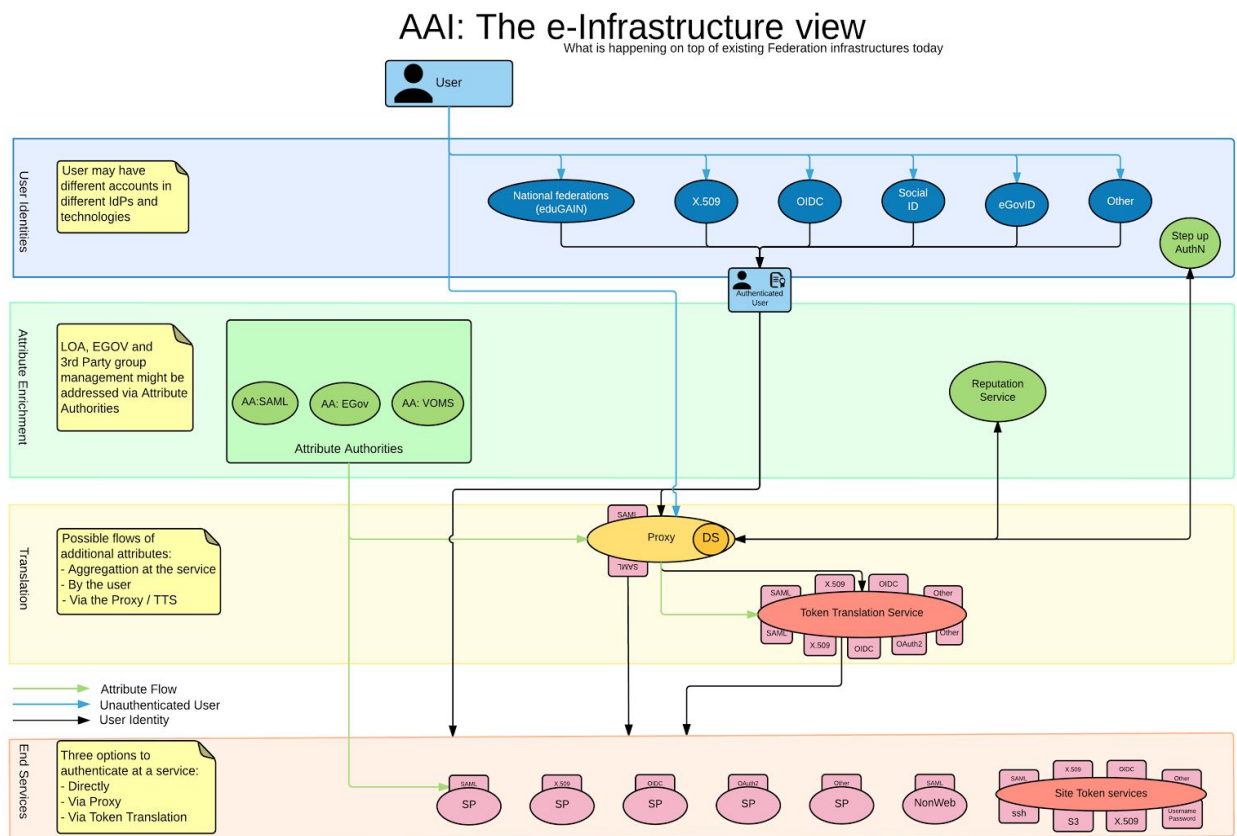

*Figure 3 - General schematic of the existing federated AAI infrastructures, with the main components placed into four layers: User Identities, Attribute Enrichment, Translation, and End Services. Arrows indicate the logical flow of information, not the workflow a user follows to access resources. Each resource (e.g. SP, IdP, AA, User) is federated in the sense that it is likely located inside a different organisational domain than other resources it interacts with.*

The diagrams in Appendix B show the information flows of the analysed architectures by highlighting the respective components. The analysis of these architectures served as input for this blueprint architecture.

In the next paragraphs we describe the functional building blocks of the blueprint architectures, to serve as an overview and to allow the identification of functional building blocks and their interplay.

The **User Identities Layer** contains services for identification and authentication of users. Usually in existing implementations in the R&E space, we find SAML[11] (Security Assertion Markup Language) Identity Providers, Certification Authorities and recently OIDC[12] (OpenID Connect) Providers (OP). Although the main focus of the services in this layer is to provide user authentication, often, certain end user profile information is released as part of the authentication process.

The **Attribute Enrichment Layer** groups services related to managing and providing information (attributes) about users. Typically, they provide additional information about the users, such as group memberships and community roles, on top of the information that might be provided by services from the User Identities Layer. Services like these exist for all the mentioned authentication technologies. Virtual Organisation Membership Service (VOMS) are commonly used in X.509 based infrastructures, Attribute Authorities (AAs) in SAML-based implementations, and the "userinfo"-endpoint in OpenID-Connect implementations. In this document we will use SAML AA terminology.

The **Translation Layer** addresses the requirement for supporting multiple authentication technologies. Most often, we encountered two types of services:

- **Token Translation Services** translate identity tokens between different technologies. Token translation can be implemented as a central service or offered at an SPs site.
- **SP-IdP-Proxy (proxy)** is an emerging pattern within research and e-infrastructures. This model is depicted in figure 2. It is predominantly found in SAML installations. Towards the Identity Federations this proxy looks like any other SP, while towards the internal SPs it acts as an IdP.

The **End Services Layer** contains the services users actually want to use. The access to these services are AAI-protected (different technologies are used to protect the access). They range from simple web services, such as wikis or portals for accessing computing and storage resources, to non-web resources like a login shell, an FTP transfer- or a workload management system.

## 3.1 Typical existing workflow

Fig. 3 shows the general AAI landscape. To illustrate a basic workflow. it is included here again.

---

[11] http://xml.coverpages.org/saml.html
[12] https://openid.net/connect/

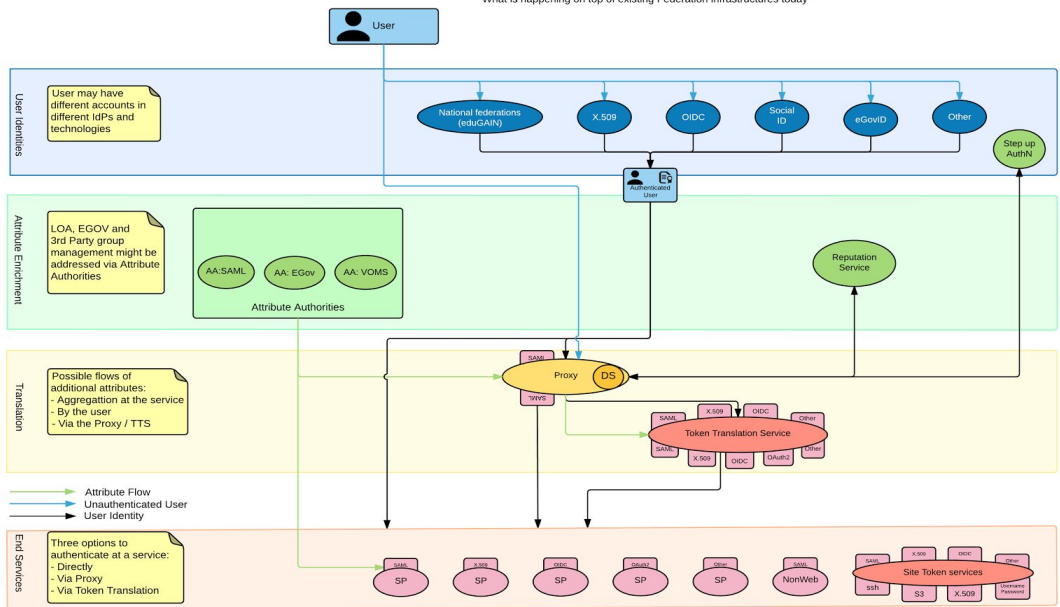

The general workflow for access to resources can be summarized:

| Ste | Technolog | Description |
|---|---|---|
| 1 | SAML OIDC | The user tries to access a resource and is redirected to an identity provider, typically via an identity provider discovery procedure. |
|   | X.509 | The user obtains a long-lived (10 days-2 years) certificate at a CA. |
| 2 | SAML | The user authenticates himself at the identity provider. The IdP creates a SAML response, which certifies his identity in the SAML-assertion and contains a set of IdP-defined attributes. The response is returned to the user. |
|   | OIDC | The user authenticates himself at the Open-ID provider. The OIDC service creates either an auth-code (pull model) or a JWT (push model). |
|   | X.509 | The user creates a proxy certificate and contacts an external Virtual Organisation Membership Service (VOMS) that certifies 3rd party attributes. |
| 3 | SAML OIDC | User is redirected back to initially contacted SP and presents the token to the SP. |
|   | X.509 | The user accesses the resource and authenticates either using the certificate retrieved in step 1 or the proxy certificate |
| 4 | SAML OIDC | The SP retrieves attributes from the token (push) and/or may use the retrieved token to retrieve additional attributes from one or more Attribute Authorities (pull). |
|   | X.509 | If a VOMS proxy certificate was used in step 3, then the SP extracts the user attributes from the proxy certificate |
| 5 | All | Based on attributes an authorisation decision is made. Access to resources is granted. |

## 3.2 Requirements covered & open challenges

In this initial draft version of the AARC Blueprint Architecture, we have worked to address a subset of the requirements coming from the user communities and infrastructure providers. In the next versions of the document we will incrementally cover more of the identified requirements. Figure 4 lists the identified requirements.

*Figure 4 - Blue: These requirements are already being addressed. Blue-Green: Work on addressing these requirements has already started. Green: These requirements are going to be addressed in the next iterations of this document.*

**Attribute Release**

The proposed architecture makes attribute release an easier case for the Identity Providers. IdPs do not have to deal with each and every service that is operated by the community. The SP-IdP Proxy provides a single point of trust between the research community and the Identity Federations and IdPs. Asserting the compliance of the policies and operational practices is much easier if it is performed for one entity instead of having to assert the compliance and establish trust with each individual community service. Furthermore, the proposed architecture allows the integration of community based and infrastructure operated Attribute Authorities, which can potentially provide many of the required user attributes. In such cases, the only thing that is required from the home IdPs is information that uniquely identifies the user in a consistent way.

**Attribute Aggregation**

Services require information that is available from multiple sources. Although the home organisation has information about the user's identity, such as name, email, and affiliation, typically information about the user's group membership(s) and role(s) within the research collaboration or research project is held by community services, which act as Attribute Authorities. Services need to be able to use both of these types of user attributes. With the proposed architecture, attribute aggregation is provided by the SP-IdP Proxy so that services will not have to implement complex technical solutions for supporting multiples AAs. Furthermore, having the attribute aggregation performed centrally by the SP-IdP Proxy allows the implementation of central strategies for attribute mapping and harmonization so that the services receive a uniform set of attributes following well defined structure and semantics.

**Community based authorisation**

The integration of external AAs, along with the ability to aggregate attributes from multiple AAs, give the ability to the communities to manage group membership and user role information internally within their collaboration. These community based user attributes can be collected by the SP-IdP proxy, which creates the "composite" user identity, combining them with the attributes retrieved from the user's home organisation. This "composite" user identity can then be made available to internal services in order to make the appropriate authorisation decisions. Still, there is a lot more work to be done in this area. With the current architecture, services can make authorisation decisions based on the user's identity. Future work will introduce the ability to manage centrally some or all of the authorisation policies, so that individual services will not have to deal with the implementation and maintenance of the authorisation policies.

**Persistent Unique Identifiers**

With the introduction of the SP-IdP proxy component, it is now possible for research communities to have one persistent, non-reassignable, non-targeted, unique identifier for each user even if such identifiers are not available from IdPs at the home organisations of the users. The initial results of the investigation[13] we conducted during the last year, have shown that the most appropriate identifier in the eduPerson schema is the eduPersonUniqueId[14] attribute. Although this is an identifier that is not (yet) available from the home organisation IdPs, it meets all of the identified requirements, and can be easily derived algorithmically based on the information available from the home organisations. Furthermore, the fact that this attribute can be scoped, is a very important characteristic for creating globally unique identifiers that can be used across infrastructures. There is still more work to be done on this subject, especially in the area of cross-infrastructure/sector interoperability. Another aspect that we are already looking into is the privacy implications of such global identifiers and how can they be addressed at the policy and technical levels.

**Guest Users**

One of the major goals of the AARC Blueprint Architecture is to also support those use cases in which users in research collaborations do not have a federated identity via their home organisation. Moreover, there are cases in which an individual researcher is not affiliated with any of the traditional home organisations. To cater for these cases, the proposed architecture enables research communities and infrastructure providers to connect to Identity Providers that are not part of any of the eduGAIN participating federations. Such Identity Providers might be special purpose Identity Providers, designated specifically for these types of users, or they can be general purpose Identity Providers such as governmental eID providers or social networks, which can act as the third party Identity Providers etc. The key factors in enabling such Guest Identity services is to be able to support multiple technologies and flexible policies in a scalable and trustworthy manner. More information about the work on Guest Identities is available in the MJRA1.2 "Design for Deploying Solutions for "Guest Identities"[15].

---

[13]https://wiki.geant.org/display/AARC/Requirements+Discussion+for+the+Blueprint+Architecture#RequirementsDiscussionfortheBlueprintArchitecture-A.2
[14]http://software.internet2.edu/eduperson/internet2-mace-dir-eduperson-201310.html#eduPersonUniqueId

**Social & e-Gov IDs**
The support for social and e-Gov IDs relates a great deal with the previous requirement for "Guest Users". The current architecture can support the use and mix of social IDs and e-Gov IDs together with identities coming from home organisations, which is partially useful for supporting the guest users. Another use cause that has been associated with the use of eGov IDs, is the support for higher levels of assurance and stronger authentication, which is also supported by the proposed architecture. Future work in this area will focus on the possibility of utilising e-Gov ID and social ID for specific use cases that require "step up authentication", instead of using them as user's primary identity.

**Levels of Assurance**
AARC has already delivered a baseline assurance framework (see MNA3.1[16]) to enable a large number of research use-cases in the infrastructures, as well as a set of differentiated assurance profiles permitting fine-grained control. However, additional assurance profiles are required to promote interoperation between infrastructures through harmonisation of LoAs or policy mappings between diverse LoAs. A framework for expressing LoA per attribute and per issuing party is missing at the present time, but is currently being investigated within AARC JRA1 and NA3 activities.

**Step up authentication**
Step up authentication will be covered in the next iterations of the AARC Blueprint Architecture

**User Friendliness**
A very important aspect of any solution is that it is user friendly and provides a good user experience. Although this is typically a responsibility of the implementers, it is also important that the architecture support user friendly implementations. Along these lines, the support of features such as multi-protocol federated access that allows the integration of a wide range of identity providers, the support for both web based and non web based workflows, etc, have a clear impact in enabling the implementation of user friendly solutions. By working together with implementers and architects who have adopted the AARC Blueprint Architecture, we have already identified some difficulties that do affect the user experience. For example, redirecting the user from one WAYF page to another when one or more IdP-Proxies are involved in the login workflow. Also, in a VO/research collaboration scenario, the user may need to explicitly select to log in through their collaboration IdP-Proxy rather than directly authenticating at their

---

[15]https://aarc-project.eu/wp-content/uploads/2016/06/MJRA1.2-Design-for-Deploying-Solutions-for-Guest-Identities.pdf
[16] https://aarc-project.eu/wp-content/uploads/2015/11/MNA31-Minimum-LoA-level.pdf

Home IdP. This may be required, for instance, when the IdP-Proxy is needed to get VO/group membership/role information required for the authorisation purposes. In the upcoming iterations of the AARC Blueprint Architecture we will be introducing implementation best practices that can be utilised for addressing such issues.

**Credential Delegation**

The proposed architecture already supports credential delegation. In the case of SAML, this can be accomplished using the Enhanced Client or Proxy (ECP) profile, however the required interaction flows are considered too complicated. At the same time, there is little (if any) support for the ECP profile in SAML implementations other than Shibboleth[17]. Due to the limited client support for the ECP profile, implementors should resort to using TTS or protocols like OIDC/OAuth2, which make simple delegation much easier. However, there are more advanced use cases where the "client" accessing the service is acting on behalf of another client that has authenticated earlier. This form of multi-tier delegation is not fully covered by the current OAuth2 specification.

**Non-web browser**

In the basic workflow for providing access to the resources (refer to Section 3.3), SAML and OIDC make use of HTTP redirects and thereby assume that the user is represented by a web browser. However, the proposed architecture also caters for non-web-based federated access to resources, such as SSH, FTP, etc. Non-web access is also required for programmatic (REST-API) access to services. The underlying requirement here is to enable seamless access without modifying the services and clients that use it. Currently, this is usually achieved via the use of certificates that can be supported through Token Translation services. Alternatives are token-based approaches, which exist for selected OAuth flows and OIDC but not for SAML. While using TTS does provide non-web access for some use cases, end-to-end access to services that does not involve a web browser is still an open issue. Non-web resources usually require local, non-transient (or ephemeral) identities, such as operating system, database or service (e.g. iRODS) user accounts. Such identities typically require prior provisioning and deprovisioning once they are no longer used. During sign-on, the federated identity of the user must be mapped to a local one. At the same time, the federated attributes must be mapped to local privileges which are typically translated to group membership or roles bindings.

**SP Friendliness**

The design of the proposed architecture has put significant focus on the friendliness towards service providers. In the proposed architecture, internal services need to trust only one single Identity Provider, the IdP subcomponent of the SP-IdP Proxy. Common technical services, which are typically required in federated access scenarios (such as Discovery Services, User Consent etc) should be provided centrally and should not have to be implemented by each service individually. Furthermore, the introduction of the IdP-Proxy attribute simplifies release, since connected SPs do not have to negotiate/agree on the release policy with each and every identity provider operated by the communities. Instead, connected SPs only deal with a single entity, i.e. the IdP-Proxy, both on the technical and policy level. Furthermore, attribute aggregation is performed by the SP-IdP Proxy so that services will not have to implement complex technical solutions for supporting multiples AAs. As such, connected SPs receive consistent/harmonised user identifiers and accompanying attribute sets that can be interpreted in a uniform way for authorisation purposes.

---

[17] https://wiki.shibboleth.net/confluence/display/CONCEPT/ECP

# 4 Summary

This document presented a general blueprint AAI architecture that can serve as a reference for the design and implementation of integrated and interoperable AAI solutions in the R&E sector. The proposed blueprint architecture builds on the current AAI landscape and is driven by the user community and service provider requirements identified at the beginning of the AARC project. As such, it aims to be integrated and to interoperate with multiple existing AAIs and different technologies. The components that comprise the architecture can be grouped into four layers, namely User Identities, Attribute Enrichment, Translation, and End Services. The various interactions/interdependencies among these components are illustrated in a high-level architectural diagram. A general workflow for giving access to the resources based on the proposed approach is also provided. The proposed architecture already addresses a number of the requirements coming from the user communities and infrastructure providers. More of the remaining challenges will be incrementally covered in the next iterations of the blueprint architecture. An overview of the analysed AAI architectures and the identified user requirements is provided in the appendix.

# APPENDIX

## Appendix A: Requirements

In its first year, the AARC project collected requirements from several scientific communities and project stakeholders. It is an outcome of the "Requirement Analysis" activity in the JRA1 work package, "Architecture for an integrated and interoperable AAI". The full report[18] is available as a PDF, a summarised list of the requirements is available in the AARC wiki pages[19]. A technical discussion of these requirements is also available in the AARC wiki pages[20].

## Appendix B: Analysed Architectures

All architecture diagrams were taken unmodified from the projects that developed them. Therefore, they expose different levels of detail. To identify the building blocks the following color coding is used:

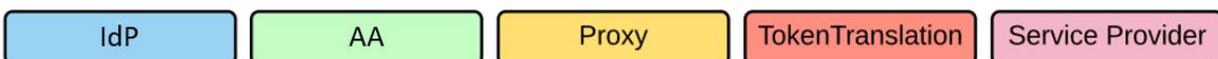

---

[18] https://aarc-project.eu/wp-content/uploads/2015/10/AARC-DJRA1.1.pdf
[19] https://wiki.geant.org/display/AARC/Collected+Requirements
[20] https://wiki.geant.org/display/AARC/Requirements+Discussion+for+the+Blueprint+Architecture

## B.1 EUDAT

EUDAT2020 is a Horizon 2020 project building a "collaborative data infrastructure," *i.e.* an infrastructure which offers data services to researchers. It offers services for replication (B2SAFE), sharing (B2SHARE), personal storage (B2DROP), moving to and from other infrastructures such as PRACE or EGI (B2STAGE), and more. The initial user communities were linguistics (CLARIN), climate (IS-ENES), Earth observation (EPOS), and human physiology (VPH), but the communities have since expanded beyond this.

The EUDAT infrastructure is implementing an AAI solution called B2ACCESS. It serves as a central proxy component which bridges various AAI technologies used within the user communities (SAML/eduGAIN, OpenID Connect) to various technologies used by the service providers within the infrastructure. B2ACCESS also manages additional attributes for the users, some of which are provided by the home IdP, some are self-asserted by the user or allocated by a group manager. The collected attributes are passed on (pushed) to the services in the created token or made available (pull): the token translation component creates the tokens required for accessing a specific service. Supported are X.509 (push), OAuth2/OIDC (pull), and SAML (push), as well as an API for account synchronisation with services (pull).

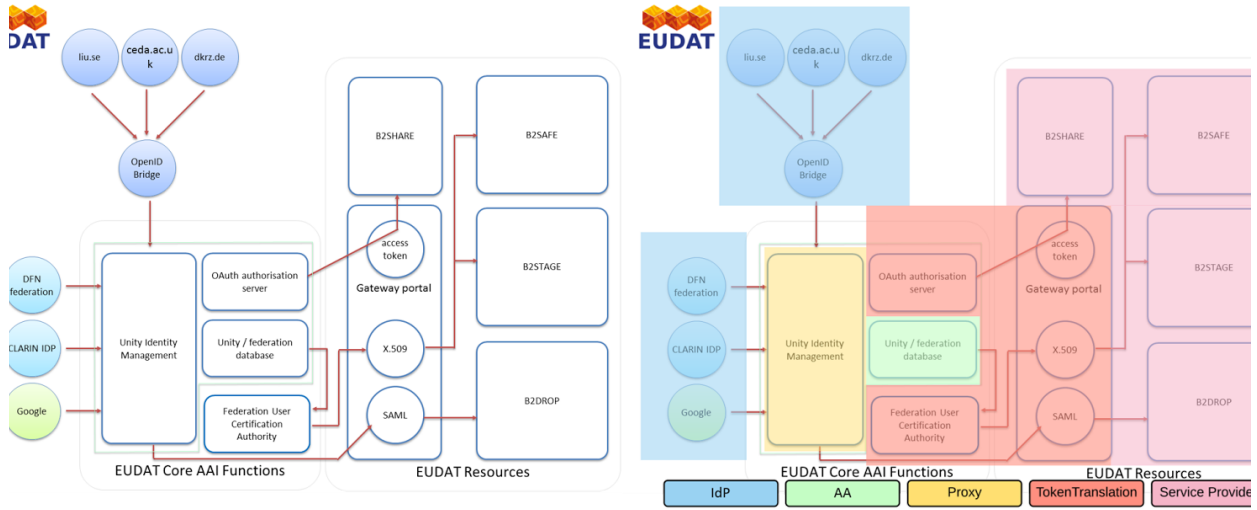

*Figure 3.1 EUDAT components mapped to color coded architectural functional components. The architectural drawing is provided by EUDAT, the color coding by AARC*

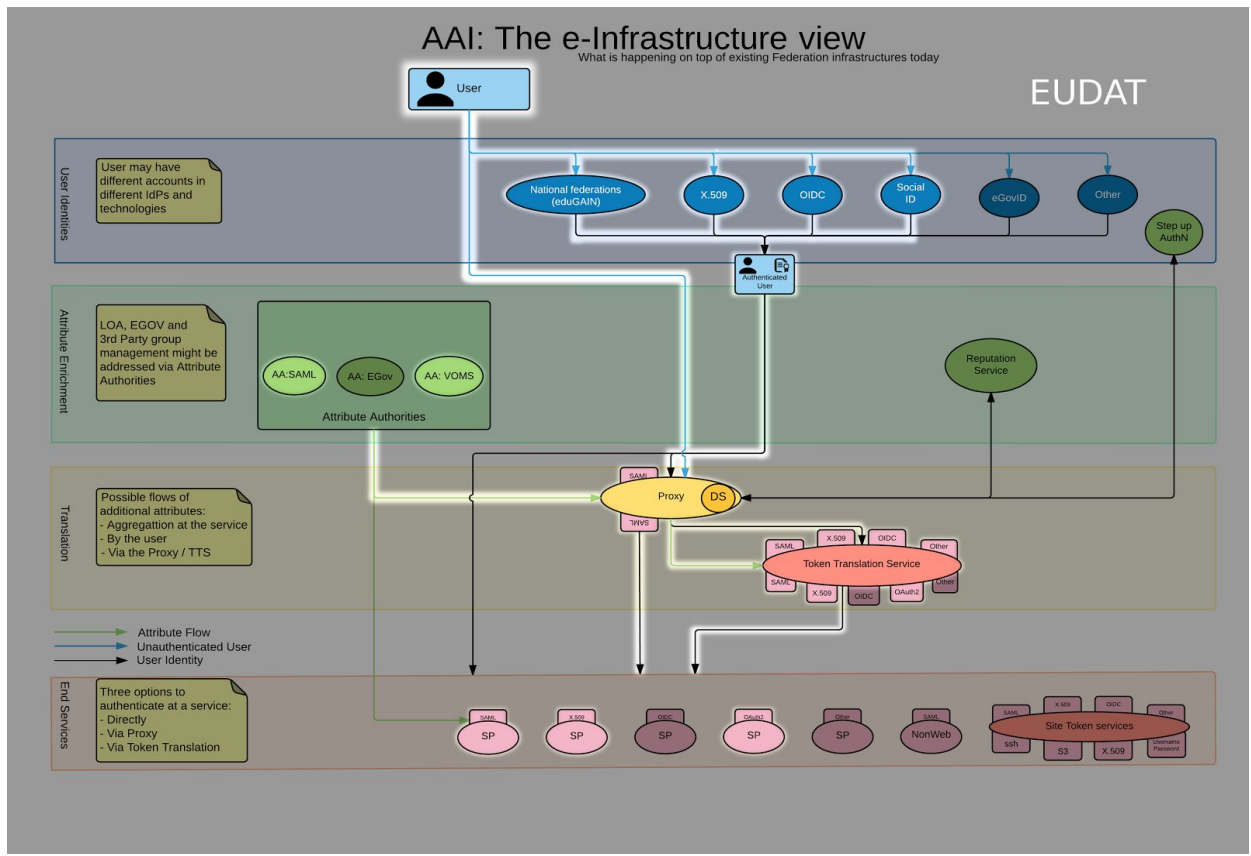

*Figure 3.2 EUDAT Authentication flow mapped within the general AAI schematic*

## B.2 DARIAH

The Digital Research Infrastructure for the Arts and Humanities (DARIAH[21]) aims to facilitate long-term access to, and use of, all European Arts and Humanities digital research data. It supports digital research, as well as the teaching of digital research methods. The DARIAH AAI is based on SAML. DARIAH provides an architecture to facilitate attribute aggregation at the SP without the need of a central proxy. They use an additional redirect to the DARIAH registration SP. This enables accepting users from IdPs that do not release enough attributes for an SP to provide a service. All these attributes are stored in the DARIAH Attribute Authority (AA).

---

[21] https://dariah.eu/

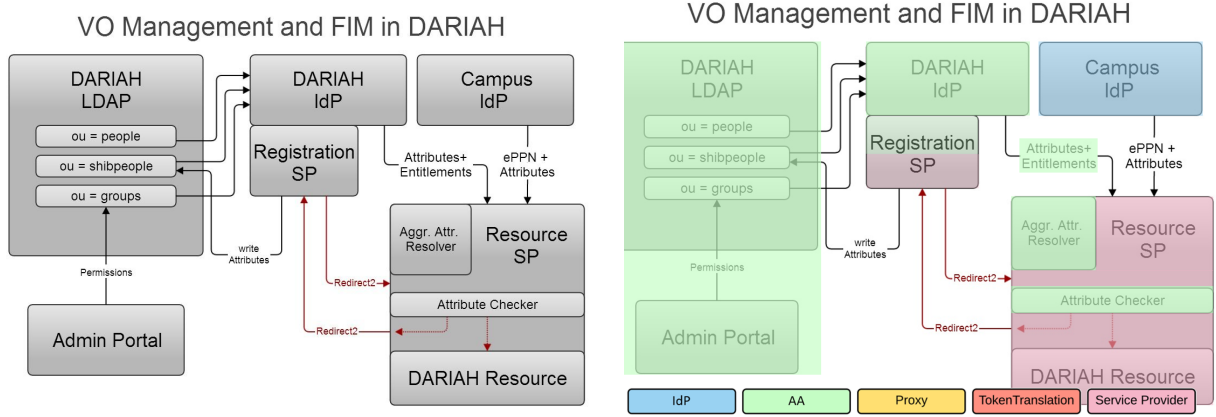

*Figure 3.3 DARIAH components mapped to color coded architectural functional components. The architectural drawing is provided by DARIAH, the color coding by AARC*

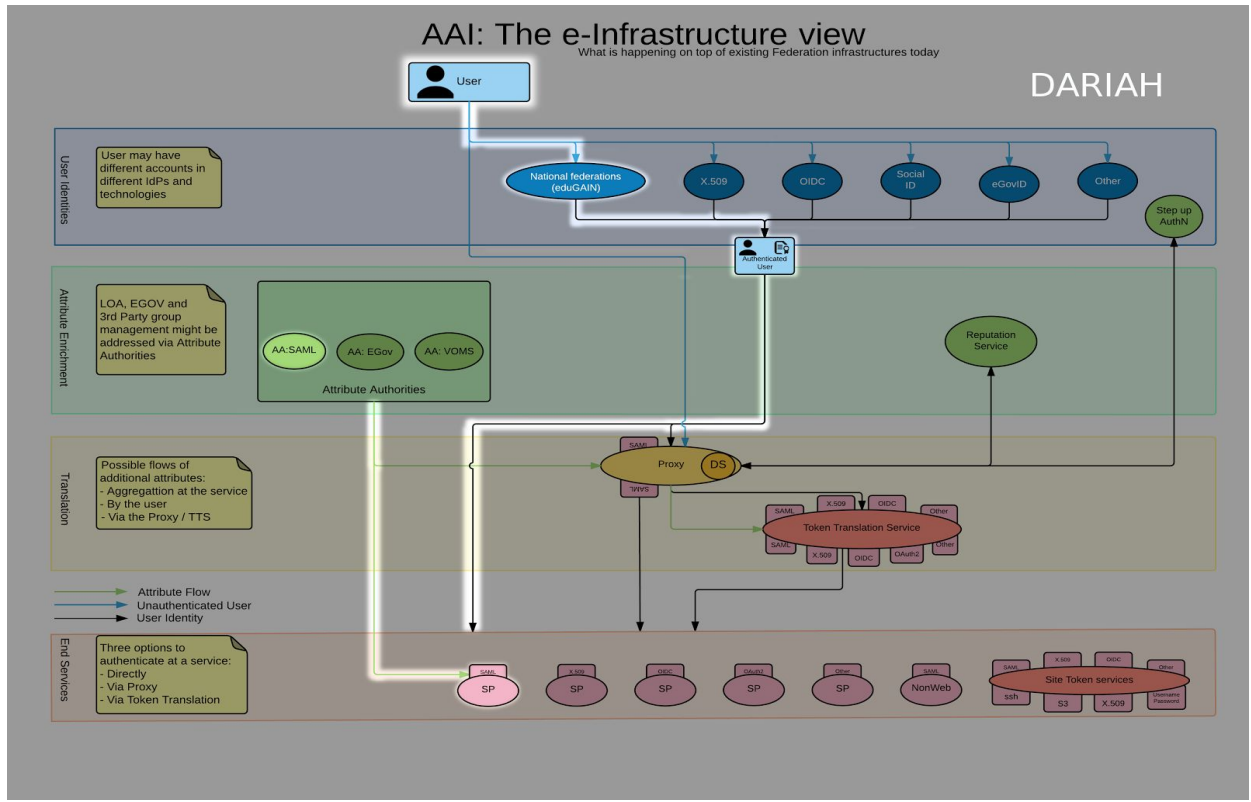

*Figure 3.4 DARIAH Authentication flow, mapped within the general AAI*

## B.3. ELIXIR

ELIXIR[22] is creating an infrastructure that integrates large amounts of biological data produced by life science experiments across Europe. From a high-level functional point of view, the ELIXIR AAI is very similar to that of EUDAT. However, the used components are different ones. ELIXIR users can sign into the ELIXIR Proxy IdP using different authentication technologies (SAML2, OIDC, eGov, …) using existing authentication systems like eduGAIN or a social IDs (Google, LinkedIn, ORCID, ...). Every user gets an ELIXIR ID, which is then used as a persistent identifier at ELIXIR services. The ELIXIR Proxy IdP collects attributes from external providers and from the ELIXIR Directory. It either authenticates the user towards a SAML SP or towards a "Credential Translation" component. This allows support for the required authentication technologies on the service provider side.

Notably, the ELIXIR Proxy IdP will support "Step-up-Authentication," where authenticated users can be authenticated again, to improve the confidence an SP may have in their identity. Improving trust in the user's identity can also be done by other methods (reputation, voting by existing members, account linking, …).

For authorisation, in the life sciences, there are services (such as access to samples donated for research by patients) that require a researcher to present a research plan to a data access committee. Dataset authorisation management is a service that facilitates that workflow. For less sensitive services, requiring only lightweight access control, there is a concept of "bona fide researcher" (a researcher in a good standing) that is enough to qualify him/her to access the service.

---

[22] https://www.elixir-europe.org/

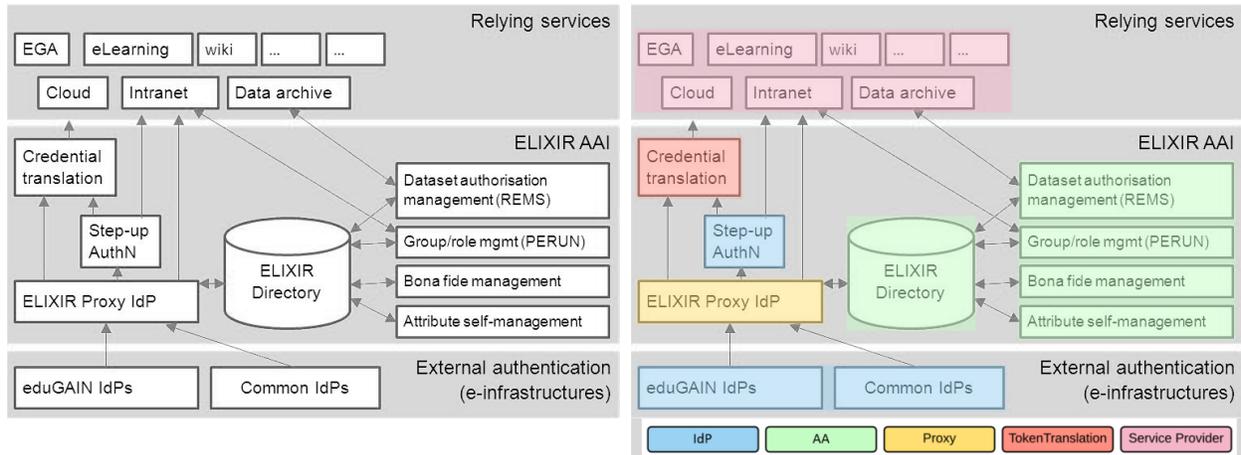

*Figure 3.5 ELIXIR components mapped to color coded architectural functional components. The architectural drawing is provided by DARIAH, the color coding by AARC*

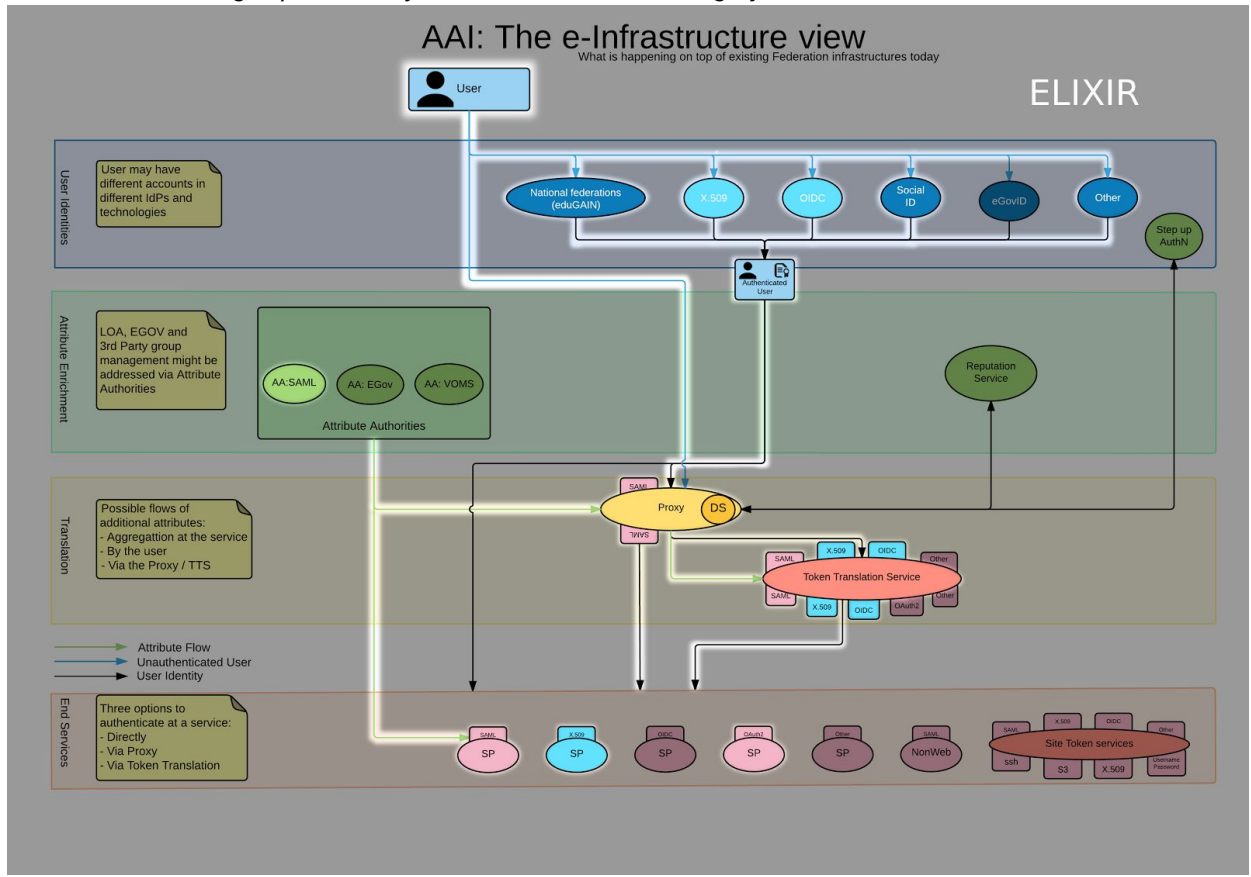

*Figure 3.6 ELIXIR Authentication flow*

## B.4 WLCG Federated Grid-Service Access

WLCG[23] is an infrastructure centred at CERN in which authentication is governed by the ownership of personal X.509 certificates. A large number of WLCG Grid Services run on this infrastructure and rely on these certificates as well as authorisation roles defined in the VOMS service to validate users. There is currently no access to this infrastructure via eduGAIN. Moving away from X.509 certificates and towards federated login via SAML is a strategic decision.

FTS ("File Transfer Service") is a service for users to schedule file transfers. It is one of many services that run on WLCG and was chosen for the initial pilot due to its suitability for enhancement and appropriate level of complexity. The FTS service itself uses X.509 certificates (or proxies[24]) to transfer files from one endpoint to another and supports a range of endpoint protocols used by grids and clouds, including GridFTP[25], WebDAV[26], and SRM[27].

WebFTS is a web (portal) front end to FTS, enabling users to easily register and monitor file transfers. WebFTS, for the ATLAS VO, was chosen as a pilot. The user authenticates via eduGAIN to the CERN SSO (as a central proxy component). The portal then transforms the SAML token into a short lived X.509 credential via STS (Security Token Service) for the user. In this pilot, participating sites had to trust the IOTA CA.

---

[23] http://wlcg.web.cern.ch/
[24] http://www.rfc-editor.org/rfc/rfc3820.txt
[25] http://www.ogf.org/documents/GFD.47.pdf
[26] http://www.webdav.org/specs/
[27] http://www.ogf.org/documents/GFD.129.pdf

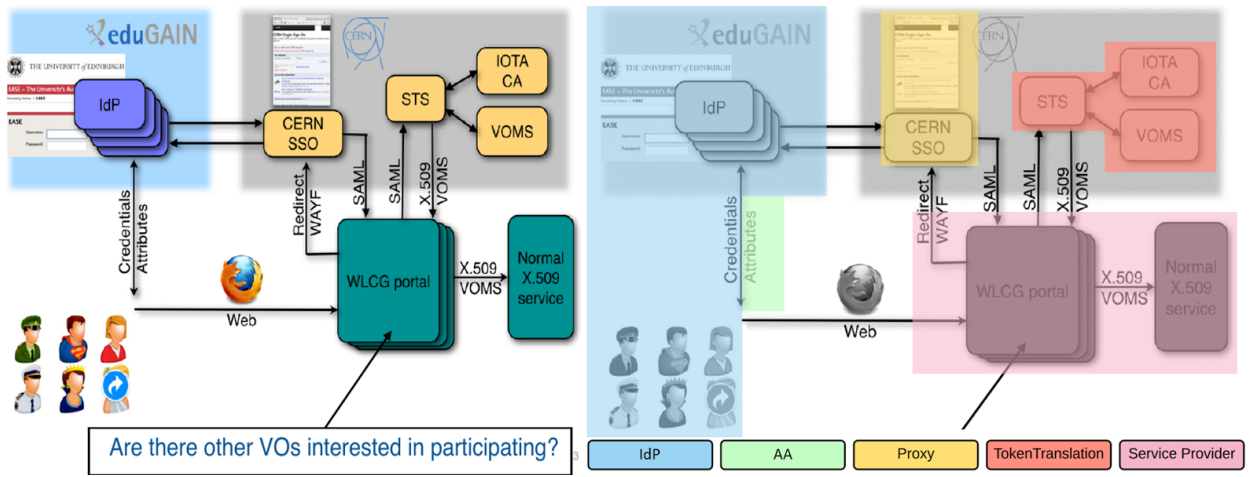

*Figure 3.7 WLCG-Federated Grid components mapped to color coded architectural functional components. The architectural drawing is provided by WLCG, the color coding by AARC*

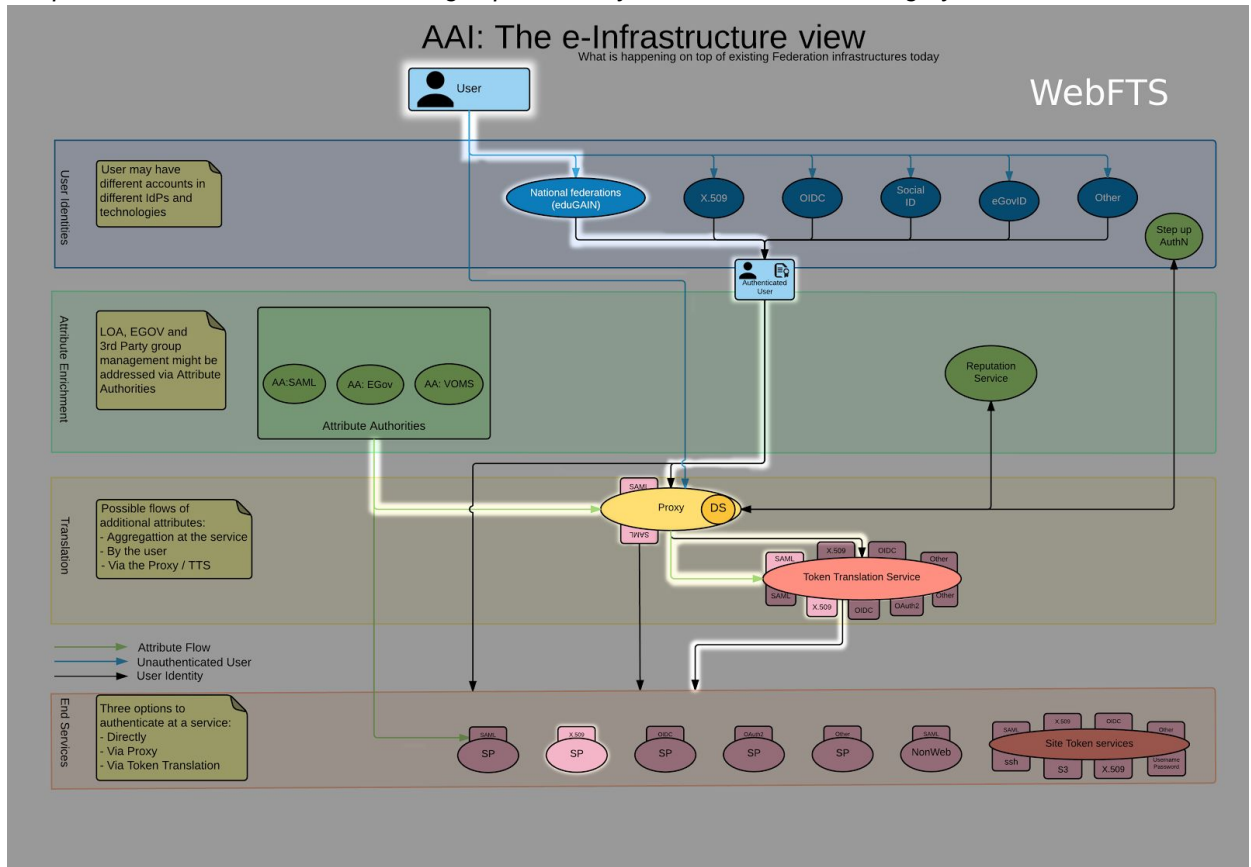

*Figure 3.8 WLCG Authentication flow*

## B.5 Umbrella

Umbrella is the AAI developed by PaNdata (Photon and Neutron Data Infrastructure) for the European Neutron and Photon community.

Umbrella defines its own SAML-based federation. It is not associated with any national identity federation or eduGAIN. Umbrella provides one global, replicated IdP that provides a persistent and unique user identifier over the lifetime of a user. Users can sign up in a self-service fashion for an account in this IdP. Authentication information is provided and updated by the users themselves (self assertion). The Umbrella account needs to be linked to a trusted one at each facility. This avoids central handling of a complex general model for trust relations and procedures between participating facilities (sites), but multiplies work for user ID vetting by delegating this to each facility.

Due to the initial low level of assurance and due to the nature of workflows in the targeted lightsource community (users travel to facilities), their identity has to be vetted upon initial visit at each facility, e.g. by providing a photo-ID for verification. Login to the Umbrella SP-IdP-proxy using a home identity is on their roadmap.

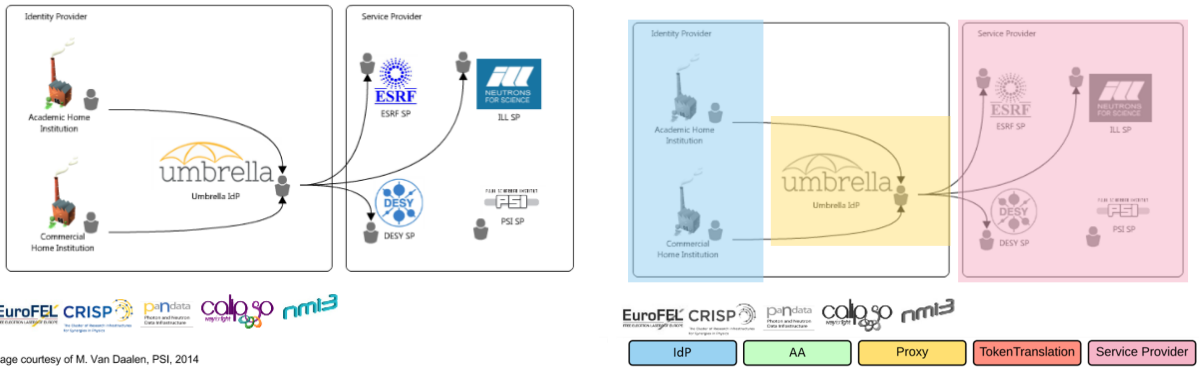

*Figure 3.9 UMBRELLA components mapped to color coded architectural functional components. The architectural drawing is provided by UMBRELLA, the color coding by AARC*

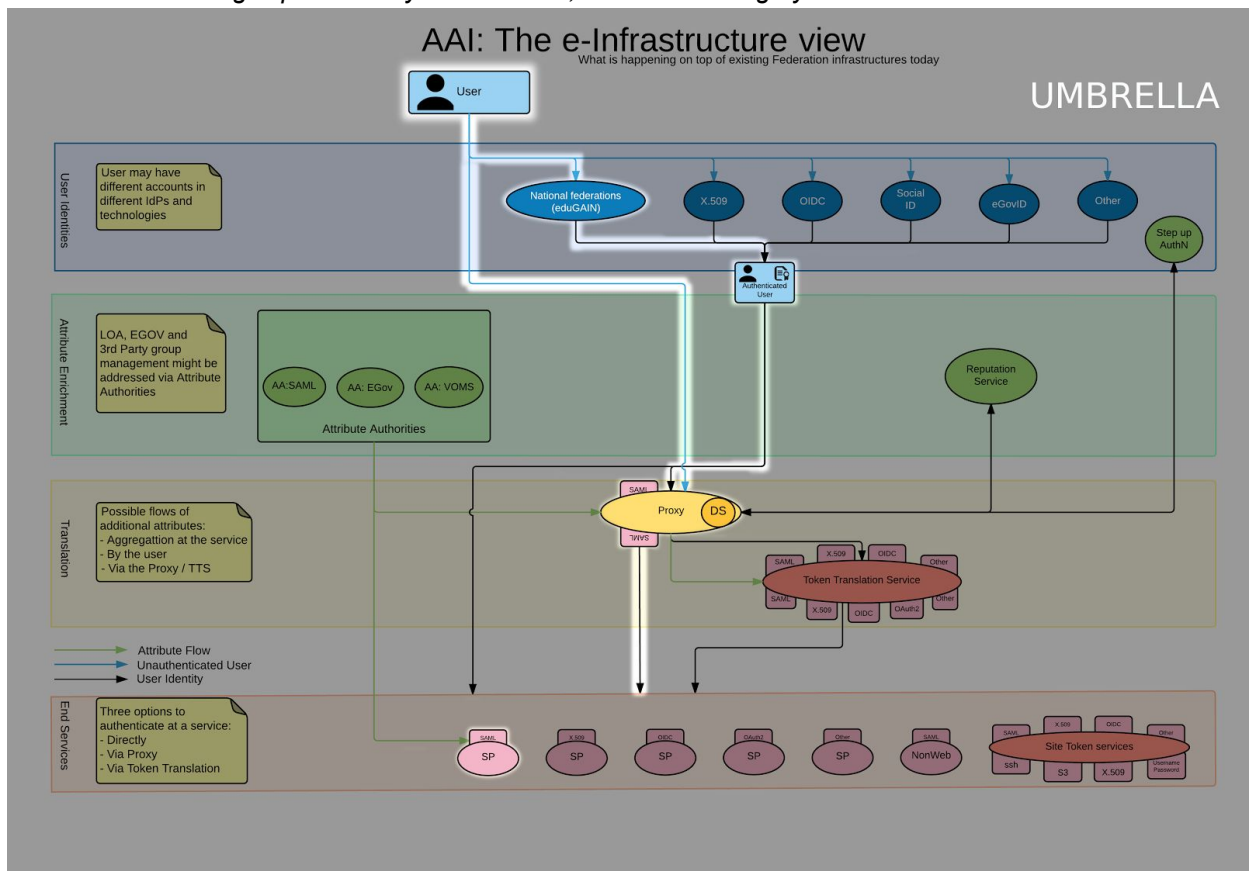

*Figure 3.10 UMBRELLA Authentication flow*

## B.6 INDIGO

INDIGO-DataCloud is an EU project, started in April 2015 and will end in September 2017. The main objective is to develop an Open-Source platform for scientific computing and data storage, deployed on public and private cloud infrastructures.

The INDIGO AAI is currently in its early implementation phase. It is going to be similar in design to the EUDAT and ELIXIR AAIs in that it allows logins from home-IdPs and social IdPs (Google, etc), and provides a proxy component. The difference however, is that credential translation occurs in two places. The first translation at the proxy always translates to OpenID-Connect (OIDC). An optional second translation occurs at each site that supports INDIGO services. That translation step can generate "whatever it takes" for a user to login. A user authenticated on a higher level (e.g. INDIGO Login Service, *proxy* in the picture below) could for example request the creation of a user account together with a site-trusted proxy-certificate, and use these for native access to X.509 based services at that site. This way INDIGO will enable users to authenticate to existing, unmodified services. Client software will be developed to hide that complexity from the user.

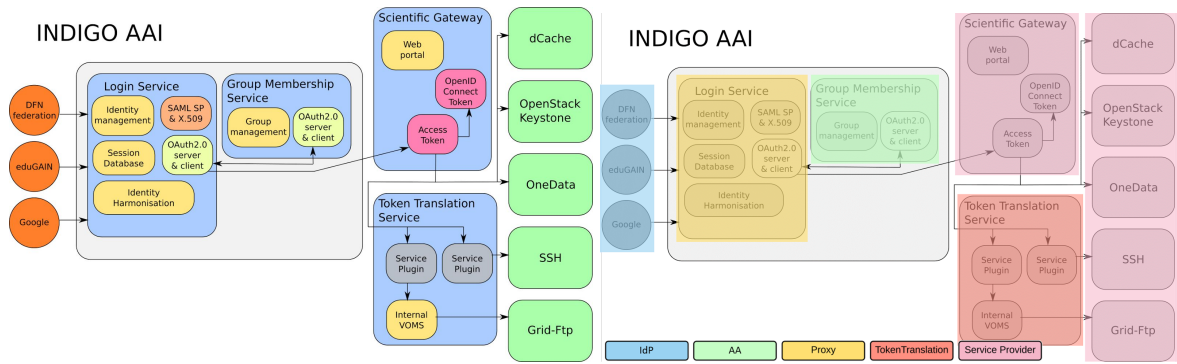

*Figure 3.11 INDIGO components mapped to color coded architectural functional components. The architectural drawing is provided by INDIGO, the color coding by AARC*

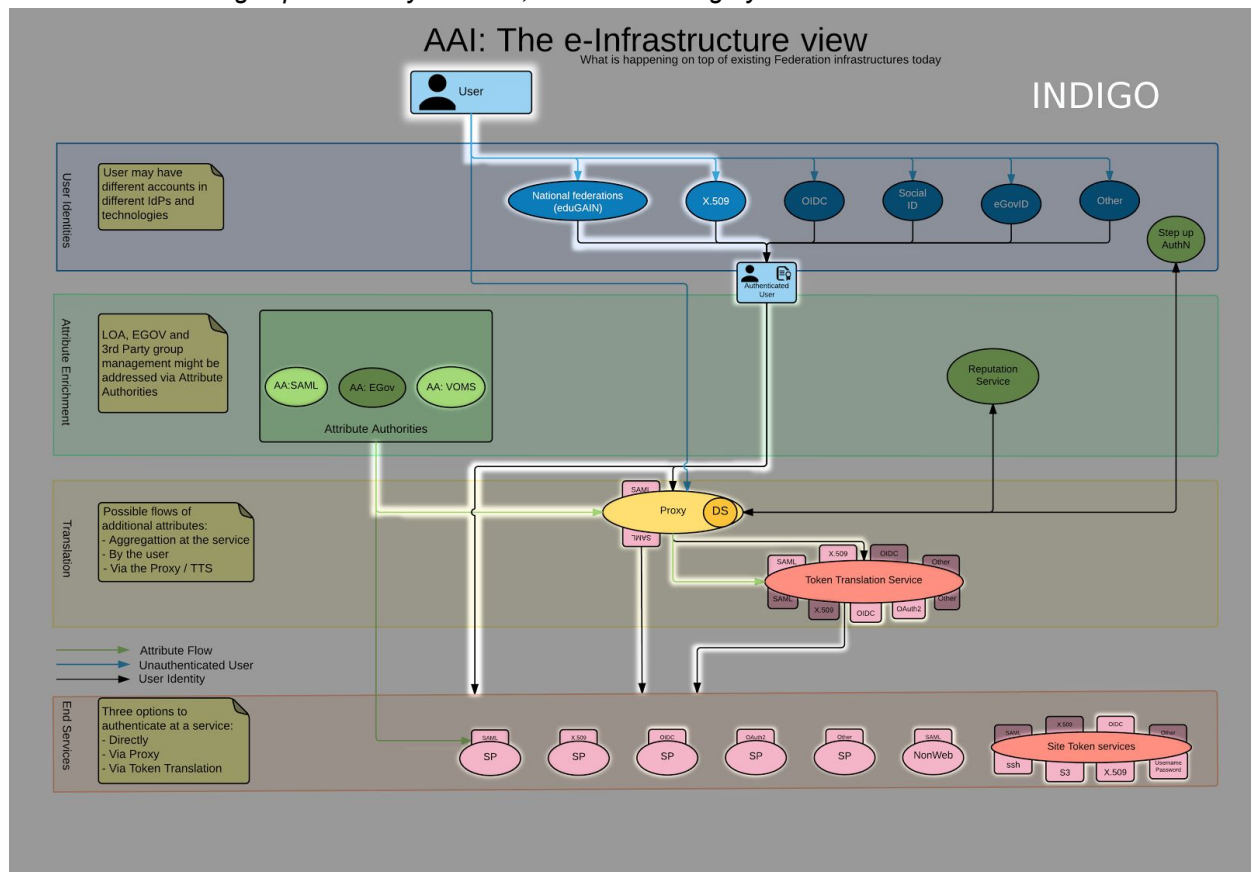

*Figure 3.12 INDIGO Authentication flow*

## B.7 EGI-AAI Pilot

The EGI AAI pilot aims to enable users to access EGI services using home IdP credentials via eduGAIN. This e-Infrastructure pilot defines an SP-IdP-proxy that allows login from eduGAIN, social identity providers (e.g. google) or using an eGOV ID. The SP-IdP-proxy uses one or more attribute authorities to extend the available attributes, and uses a token translation service to generate X.509 certificates. These certificates can be used to authenticate at all services provided within the EGI infrastructure.

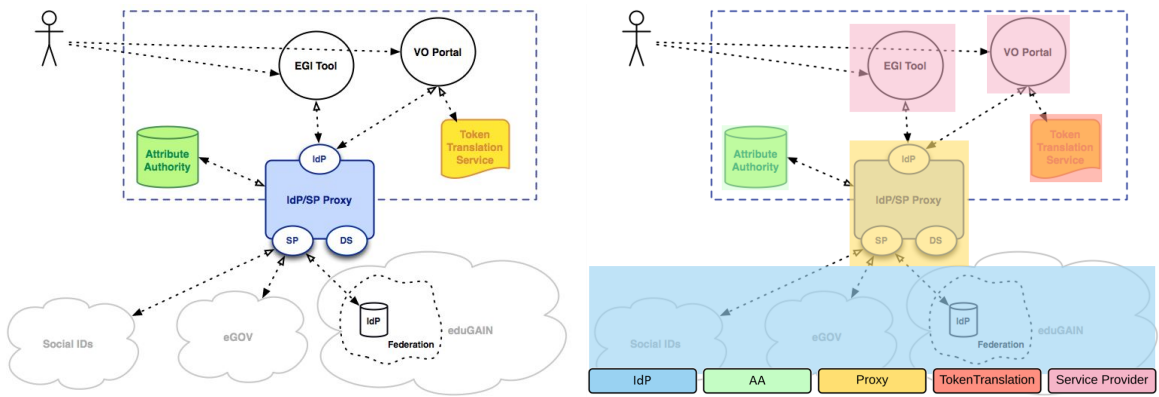

*Figure 3.13 EGI components mapped to color coded architectural functional components. The architectural drawing is provided by EGI, the color coding by AARC*

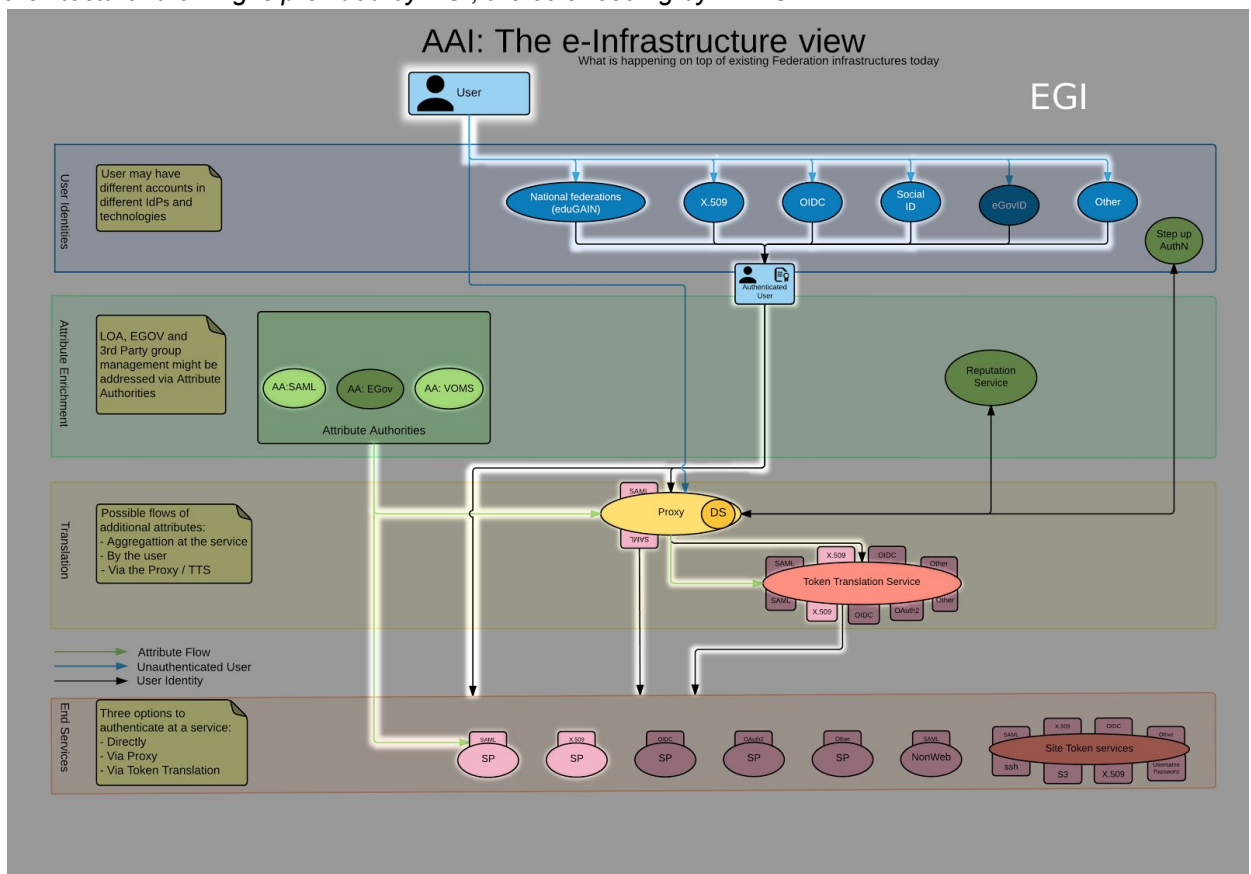

*Figure 3.14 EGI Authentication flow*